\documentclass[Journal,NoLineNumbers,letterpaper]{ascelike-new}

\usepackage[utf8x]{inputenc}
\usepackage[T1]{fontenc}
\usepackage{lmodern}
\usepackage{graphicx}
\usepackage[figurename=Fig.,labelfont=bf,labelsep=period]{caption}
\usepackage{subcaption}
\usepackage{amsmath}
\usepackage[colorlinks=true,citecolor=red,linkcolor=black]{hyperref}
\usepackage{scalerel}
\usepackage{dirtytalk}
\DeclareMathOperator*{\argminA}{arg\,min} 

\usepackage{graphicx}
\usepackage{scalerel}
\usepackage{epstopdf}
\usepackage{amssymb}
\usepackage{bm}
\usepackage{float}
\usepackage{paralist}
\usepackage{booktabs}
\usepackage{amsfonts}
\usepackage[usenames]{color}
\usepackage{array}
\usepackage{mathrsfs}
\usepackage{mathtools}
\usepackage{amsthm}
\usepackage{epic,eepic,pstricks}
\usepackage{rotating}
\usepackage{tikz}

\NameTag{Roohi, \today}

\begin{document}

	\title{Reconstructing Element-by-Element Dissipated Hysteretic Energy in Instrumented Buildings: Application to the Van Nuys Hotel Testbed}
	
	\author[1,*]{Milad Roohi}
	\author[2]{Eric M. Hernandez}
	\author[3]{David Rosowsky}
	
	\affil[1]{Postdoctoral fellow, NIST Center of Excellence for Risk-Based Community Resilience Planning, Department of Civil and Environmental Engineering, Colorado State University, Fort Collins, CO 80523, USA; formerly, Graduate Research Assistant, Department of Civil and Environmental Engineering, University of Vermont, Burlington, VT 05405 USA. E-mail: mroohigh@colostate.edu}
	\affil[2]{Gregory N. Sweeny Associate Professor of Civil Engineering, Department of Civil and Environmental Engineering, University of Vermont, Burlington, VT 05405 USA. E-mail: eric.hernandez@uvm.edu}
	\affil[2]{Professor of Civil Engineering, Former Provost and Senior Vice President, University of Vermont, Burlington, VT 05405 USA. E-mail: david.rosowsky@uvm.edu}
	\affil[*]{Corresponding author: Milad Roohi, mroohigh@colostate.edu}
	
	\maketitle
	\begin{abstract}
		The authors propose a seismic monitoring framework for instrumented buildings that employs dissipated energy as a feature for damage detection and localization. The proposed framework employs a nonlinear model-based state observer, which combines a nonlinear finite element model of a building and global acceleration measurements to estimate the time history of seismic response at all degrees of freedom of the model. This includes displacements, element forces, and plastic deformations in all structural members. The estimated seismic response is then used to 1) estimate inter-story drifts and determine the post-earthquake re-occupancy classification of the building based on performance-based criteria, 2) compare the estimated demands with code-based capacity and reconstruct element-by-element demand-to-capacity ratios and 3) reconstruct element-level normalized energy dissipation and ductility. The outcome of this process is employed for the performance-based monitoring, damage detection, and localization in instrumented buildings. The proposed framework is validated using data from the Van Nuys hotel testbed; a seven story reinforced concrete building instrumented by the California Strong Motion Instrumentation Program (Station 24386). The nonlinear state observer of the building is implemented using a distributed plasticity finite element model and seismic response measurements during the 1992 Big Bear and 1994 Northridge earthquakes. The performance and damage assessment results are compared with the post-earthquake damage inspection reports and photographic records. The results demonstrate the accuracy and capability of the proposed framework in the context of a real instrumented building that experienced significant localized structural damage.
	\end{abstract}
	
	\section{Introduction}
	
	This paper proposes a seismic monitoring framework that employs dissipated energy as a feature for damage detection and localization in instrumented moment resisting frame building structures. The main advantages of the proposed energy-based approach are: i) the proposed feature is physically meaningful and correlates well with the level of cyclic damage experienced during strong earthquakes \cite{Uang1990,sucuoglu2004energy,teran2007energy}, ii) dissipated energy can be reconstructed from element level stress-strain fields, which can be estimated from global acceleration measurements \cite{stephens1987damage,Roohi2019nonlinear}, and iii) it can be calibrated using experimental data \cite{krawinkler1983cumulative,park1985mechanistic,sucuoglu2004energy}. Despite the immediate appeal, the application of this feature for structural health monitoring purposes has been limited \cite{Frizzarin2010damage,hernandez2012dissipated} primarily due to the challenges associated with estimating dissipated energy under dynamic loading. The main contribution of this paper consists in reconstructing element-by-element dissipated hysteretic energy {\color{black}using a nonlinear model-data fusion approach.} This approach deviates from the traditional approach used in structural monitoring and damage identification, which seeks changes in the structural parameters before and after an earthquake.  
	
	{\color{black} To contextualize the proposed energy-based method, current damage detection methods are briefly reviewed. Based on the damage features selected to distinguish between undamaged and damaged states of the structure, the existing methods can be 
		widely categorized into 1) spectral, 2) wave propagation, 3) time series, 4) demand-to-capacity ratio, 5) model updating methods.
		
		The spectral methods assume that changes in spectral parameters (mode shapes and frequencies) of a structure indicate the occurrence of structural damage; where the changes are identified from vibration measurements before/after or during strong ground motion. The main challenges associated with this approach include: i) spectral parameters can be conceptually defined if the dynamic response is governed by a linear equation of motion; however, this feature does not exist for nonlinear hysteretic structural systems, ii) damage localization is a challenging task using changes in spectral parameters; this is mainly because low-frequency modes are the only reliable modes identified from vibration data, and the sensitivity of these modes to localized damage is low, and iii) changes in spectral parameters can be due to other factors such as environmental effects, which can negatively affect the reliability of this approach to detect structural damage.
		
		The wave propagation methods process the measured vibration data to extract travel times of seismic waves propagating through the structure and use this feature to detect changes in structural stiffness and subsequently infer structural damage. The main advantages of this approach include i) damage detection using a small number of instruments and ii) high sensitivity to localized damage. However, the resolution of damage detection depends on the number of instruments. This means that only two instruments are enough to determine if the building is damaged, and additional instruments are required to improve the resolution and specify the damaged part or floor of the structure.
		
		The time series methods employ a data-driven approach to detect damage based on mathematical models extracted solely from measured vibrations. These methods require no information from structural models and only track changes in the time history response or the identified black box input-output model coefficients. Thus, it is a difficult task to correlate these features with the damage sensitive structural quantities. This drawback makes these methods less appealing for seismic monitoring purposes.
		
		The demand-to-capacity ratio methods operate by comparing element-level force demands with capacities of any pertinent failure mode or comparing inter-story drifts with qualitative performance criteria to assess the level of damage in a particular member or story. The intention of selecting this damage feature for damage detection is to make the results similar to the way in which the buildings are designed, making them more interpretable. However, this approach has the drawback that the expected capacities are estimated based on codes and laboratory experiments, which can differ considerably from the actual capacities of structural elements because of the uncertainty in the stiffness and strength of building materials.
		
		The model updating approach updates the structural model parameters to minimize the error between the model estimates and vibration measurements. The structural damage can be identified by seeking changes in the model parameters. The main drawback of the model updating approach include i) the effectiveness of this approach depends on the model class, and it is necessary to examine the robustness to modeling error, and ii) the uniqueness problem may arise in the case of structural models where free parameter space becomes too large. Therefore, it is necessary to have prior knowledge regarding elements likely to be damaged.
		
		The proposed energy-based damage feature can overcome some of the drawbacks associated with existing methods if element-by-element energy demands can be reconstructed from global response measurement. In this paper, the authors propose a three step process: (1) implement a state observer to reconstruct the dynamic response at all degrees of freedom (DoF) of the model, (2) use the reconstructed response to estimate element-by-element forces and displacements, (3) use estimated displacement, forces, and constitutive laws to estimate element-level dissipated energy. The accuracy of this approach depends mostly on the performance of the state observer in reconstructing the dynamic response.  Researchers have successfully implemented various nonlinear state observers including the extended Kalman filter (EKF) \cite{gelb1974applied}, unscented Kalman filter (UKF) \cite{wan2000unscented}, particle filter (PF) \cite{doucet2000sequential}, and nonlinear model-based observer (NMBO) \cite{Roohi2019nonlinear} for response reconstruction in nonlinear structural systems. The EKF, UKF, and PF are computationally intensive and require the use of rather simple state-space models, which may not be capable of capturing the complexity of nonlinear structural behavior. However, the NMBO can be implemented directly as a second-order nonlinear FE model. This capability allows the NMBO to take advantage of simulation and computation using the conventional structural analysis software for the purpose of state estimation.}
	
	{\color{black}The primary aim of this paper is to address the reconstruction of element-by-element dissipated hysteretic energy for damage detection and localization in instrumented buildings.} A seismic monitoring framework is proposed that employs the NMBO to combine a nonlinear FE model with acceleration measurements for reconstructing the complete dynamic response as the vibration measurements become available for every seismic event. Then, the estimated response is processed to 1) estimate inter-story drifts and determine the post-earthquake re-occupancy classification of the building based on performance-based criteria 2) to compare the estimated demands with code-based capacity and reconstruct element-by-element demand-to-capacity ratios and 3) reconstruct element-level dissipated energy and ductility. The outcome of this process is employed for the performance-based monitoring, damage detection, and localization of instrumented buildings.
	
	{\color{black}A secondary objective of this paper is to validate the application of the NMBO for reconstruction of nonlinear response in the context of instrumented buildings that experience severe structural damage during an earthquake.} The NMBO has been successfully validated using case study of the NEESWood Capstone project, a fully-instrumented six-story wood-frame building tested in full-scale at the E-Defense shake table in Japan.  It was demonstrated that the NMBO could estimate quantities such as drifts and internal forces from a few acceleration measurements \cite{roohi2019iwshm,Roohi2019arxiv}. However, in this test, nonlinearity was limited and distributed throughout the building{\color{black}; e.g., during the maximum credible earthquake (MCE) level test corresponding to a 2\%/50-year event, the damage was limited to nonstructural elements such as gypsum wallboard, and no structural damage was reported \cite{van2010damage}.} 
	
	{\color{black}The effectiveness of the proposed energy-based damage detection and localization method is investigated} using data from Van Nuys hotel testbed; a seven story reinforced concrete (RC) building instrumented by the California Strong Motion Instrumentation (CSMIP) Program (Station 24386). The Van Nuys building was severely damaged during the 1994 Northridge earthquake, and localized damage occurred in five of the nine columns in the 4th story (between floors 4 and 5) of the south longitudinal frame. In the literature, multiple researchers have studied this building for seismic damage assessment and localization. Traditionally, the main objective has been to use acceleration measurements to identify the presence of damage and reproduce its location and intensity with respect to the visual evidence.
	
	The remainder of this paper is organized as follows. First, the nonlinear dynamic analysis of building structures is discussed, and the system and measurement models of interest are presented. This is followed by a section on dissipated energy reconstruction and nonlinear model-data fusion in instrumented buildings. Then, a section discussing the case study of Van Nuys seven-story RC building is presented.  The paper ends with a section presenting the validation of the proposed damage detection and localization methodology using seismic response measurements of the case-study building.
	
\begin{figure}[H]
		\centering
		\includegraphics[width=0.8\linewidth]{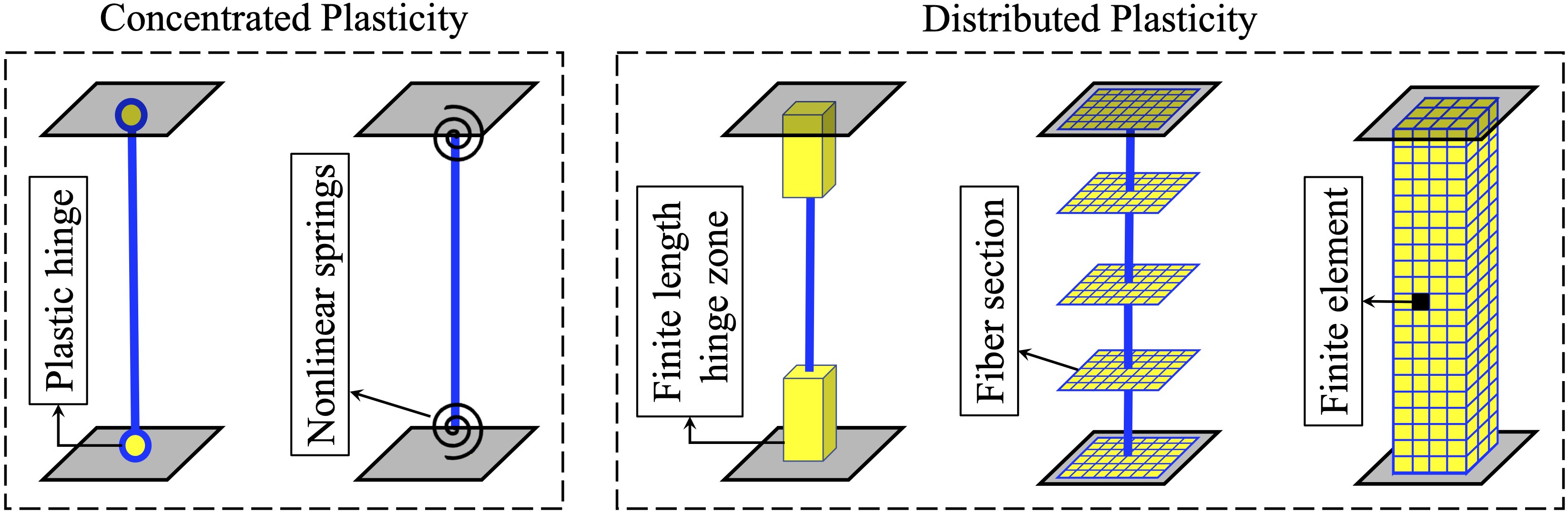}
		\vspace{-5pt}
		\caption{Schematic of nonlinear beam-column elements  (Deierlein et al. 2010)}
		\label{fig:element}
	\end{figure}

	{\color{black}\section{Structural Modeling for Nonlinear Model-Data Fusion}\label{section2}}
	Various approaches are available in the literature for the nonlinear structural modeling and dynamic analysis of moment resisting frame building structures subjected to seismic excitations. These approaches can be classified into three categories based on their scales: 1) global modeling, 2) discrete FE modeling, and 3) continuum FE modeling \cite{Taucer1991}. The \textit{global modeling} approach condenses the nonlinear behavior of a building at selected global DoF. One example is to assign the hysteretic lateral load-displacement and energy dissipation characteristics of every story of building to an equivalent element and assemble these elements to construct a simplified model of a building. This method displays low-resolution, which depending on the specific application might be detrimental.  The \textit{discrete FE modeling} approach first formulates the hysteretic behavior of elements and then, assembles interconnected frame elements to construct an FE model of a structure. Two types of element formulations are used in research and practice, including 1) a concentrated plasticity formulation and 2) a distributed plasticity formulation. The \textit{concentrated plasticity} formulation lumps the nonlinear behavior in springs or plastic hinges at the end of elastic elements. The \textit{distributed plasticity} formulation that concentrates the nonlinear behavior at selected integration points along the element using cross-sections that are discretized to fibers, which account for stress-strain relations of corresponding material. The \textit{continuum FE element modeling} approach discretizes structural elements into micro finite elements and requires localized model parameters (constitutive and geometric nonlinearity) calibration. The analysis of such high-resolution models increases the computational complexity. Therefore, this approach can be unpractical for the model-data fusion and response reconstruction applications. Figure \ref{fig:element} presents a schematic of five idealized nonlinear beam-column elements developed for nonlinear modeling of moment resisting frame building structures. 
	
	From these formulations, the concentrated and distributed plasticity formulations have been implemented in advanced structural simulation software packages such as OpenSEES, Perform, and SAP. In recent years, the fiber-based distributed plasticity FE modeling has been the most popular approach among researchers. The main reasons are: 1) the formulation accurately simulates the coupling between axial force and bending moment and also, accounts for element shear, 2) various uniaxial material models have been developed by researchers to characterize section fibers and are available for users of advanced structural simulation software, 3) the predictions using this formulation have been validated with experimental testing, and 4) the simulation and analysis are computationally efficient and accurate, even with a relatively low number of integration points per element. This paper employs a fiber-based distributed plasticity FE modeling approach for nonlinear model-data fusion and seismic response reconstruction.
	
	\subsection{System and measurement models of interest}\label{Section21}
	The global response of building structures to seismic ground motions can be accurately described as
	\begin{equation}
	\mathbf{M}\ddot{q}(t)+\mathbf{C}_{\xi}\dot{q}(t)+f_R(q(t),\dot{q}(t),z(t))=-\mathbf{M}\mathbf{b}_1\ddot{u}_g(t)+\mathbf{b}_2w(t) 
	\label{system}
	\end{equation}
	where the vector $q(t)\in \mathbb{R}^{n}$ contains the relative displacement (with respect to the ground) of all stories. $z(t)$ is a vector of auxiliary variables dealing with material nonlinearity and damage behavior. $n$ denotes the number of geometric DoF, $\mathbf{M}=\mathbf{M}^T \in  \mathbb{R}^{n \times n}$ is the mass matrix, $\mathbf{C}_{\xi}=\mathbf{C}_{\xi}^T \in \mathbb{R}^{n\times n}$ is the damping matrix, $f_R(\cdot) $ is the resultant global restoring force vector. The matrix $\mathbf{b}_1 \in  \mathbb{R}^{n \times r}$ is the influence matrix of the $r$ ground acceleration time histories defined by the vector $\ddot{u}_g(t) \in  \mathbb{R}^{r}$. The matrix $\mathbf{b}_2 \in  \mathbb{R}^{n \times p}$ defines the spatial distribution the vector $w(t) \in \mathbb{R}^p$, which in the context of this paper represents the process noise generated by unmeasured excitations and (or) modeling errors. 
	
	This study relies only on building vibrations measured horizontally in three independent and non-intersecting directions {\color{black} and assumes the vector of acceleration measurements, $\ddot{y}(t) \in \mathbb{R}^{m}$, is given by
		\begin{equation}
		\ddot{y}(t) = -\mathbf{c}_2\mathbf{M}^{-1}\left[\mathbf{C}_{\xi}\dot{q}(t)+f_R(q(t),\dot{q}(t),z(t)) -\mathbf{b_2}w(t)\right]+\nu(t)
		\label{ACC}
		\end{equation} 
		where 
		$\mathbf{c}_2 \in  \mathbb{R}^{m \times n}$ is a Boolean matrix that maps the DoFs to the measurements, and $\nu(t) \in \mathbb{R}^{m \times 1}$ is the measurement noise. 
		
		\section{Dissipated Energy Reconstruction from Response Measurements}
		This section presents the theoretical background necessary to calculate dissipated energy induced by material nonlinearity and proposes a nonlinear model-data fusion approach to reconstruct element-by-element dissipated energy from global response measurements of building structures.
		
		\subsection{Theoretical background}
		The dissipated hysteretic energy ($E_h$) can be defined by a change of variables and integrating equation of motion in time for multi-DoF systems as follows
		\begin{equation}
		\int\dot{q}(t)^T\mathbf{M}\ddot{q}(t)dt+\int\dot{q}(t)^T\mathbf{C}_{\xi}\dot{q}(t)dt+\int\dot{q}(t)^Tf_R(q(t),\dot{q}(t),z(t))dt=-\int\dot{q}(t)^T\mathbf{M}\mathbf{b}_1\ddot{u}_g(t)dt
		\label{MDOF}
		\end{equation}
		Equation \ref{MDOF} can be represented in energy-balance notation \cite{Uang1990} as follows
		\begin{equation}
		E_k+E_\xi+E_{s}=E_{i}    
		\label{energy}
		\end{equation}
		where $E_k$, $E_\xi$, $E_s$ and $E_{i}$ are kinetic, viscous damping, stain and input energy, respectively. The strain energy is the sum of recoverable elastic strain energy ($E_e$) and irrecoverable dissipated hysteretic energy ($E_h$). Thus, Equation \ref{energy} can be written as
		\begin{equation}
		E_k+E_\xi+(E_e+E_h)=E_{i}    
		\label{energy1}
		\end{equation}
		The dissipated hysteretic energy ($E_h$) can be calculated using element-level stress-strain or force-displacement demand by integrating the area under hysteresis loops as follows
		\begin{equation}
		E_h = \dfrac{1}{2}\int \epsilon^T\sigma  dV
		\label{energy0}
		\end{equation}
		where $\sigma$ and $\epsilon$ are stress and strain demands and $V$ is the total volume of an element. In distributed plasticity beam-column elements, where energy dissipation occurs primarily due to bending, the dissipated hysteretic energy ($E_h$) can be calculated by integrating the moment-curvature response along the element as follows
		\begin{eqnarray}\label{E0}
		E_h = \int_{0}^{L}M\phi dx =\sum_{j=1}^{N_p}(M\phi |_{x=\xi_j})\omega_j
		\label{energy2}
		\end{eqnarray}
		where $M$ and $\phi$ are moment and curvature response of elements, respectively; $N_p$ is number of integration points along the element; $\xi_j$ and $\omega_j$ respectiveky denote locations and associated weights of integration points. 
		
		As can be seen from Equations \ref{energy0} and \ref{energy2}, the calculation of $E_h$ requires element-level seismic response to be known. Therefore, there is a need to employ signal processing algorithms that can accurately reconstruct the element-level seismic response from global response measurements. The next subsection addresses this need by proposing the use of a recently developed nonlinear model-data fusion algorithm for seismic response reconstruction.
		
		\subsection{Nonlinear model-data fusion and seismic response reconstruction}
		Recently, \cite{Roohi2019nonlinear} proposed a nonlinear state observer for nonlinear model-data fusion in second-order nonlinear hysteretic structural systems. This nonlinear state observer has appealing properties for seismic monitoring application; two most important ones include:  (1) it has been formulated to be realizable as a nonlinear FE model, which allows implementing the nonlinear state observer using the conventional structural analysis software; therefore, the computational costs would reduce significantly, and (2) it uses power spectral density (PSD) representation to account for measurement noise and unmeasured excitations explicitly. This property is important as it is consistent with the representation of seismic excitation in many stochastic models.
		
		The NMBO estimate of the displacement response, ${\hat{q}}(t)$, is given by the solution of the following set of ordinary differential equations
		\begin{eqnarray}\label{NMBO}
		\mathbf{M}\ddot{\hat{q}}(t)+(\mathbf{C}_{\xi}+\mathbf{c}_{2}^{T}\mathbf{E}\mathbf{c}_{2})\dot{\hat{q}}(t)+f_R(\hat{q}(t),\dot{\hat{q}}(t),z(t))=\mathbf{c}_{2}^{T}\mathbf{E}\dot{y}(t)
		\end{eqnarray}
		{\color{black}where $\dot{y}(t)$ is the measured velocity and $\mathbf{E}\in  \mathbb{R}^{m \times m}$ is the feedback gain. It can be seen that Equation \ref{NMBO} is of the same form of the original nonlinear model of interest in Equation \ref{system}. A physical interpretation of the NMBO can be obtained by viewing the right-hand side of Equation \ref{NMBO} as a set of corrective forces applied to a modified version of the original nonlinear model of interest in the left-hand side. The modification consists in adding the damping term $c_2^T\mathbf{E}c_2$, where the matrix $\mathbf{E}$ is free to be selected. The diagonal terms of $\mathbf{E}$ are equivalent to grounded dampers in the measurement locations, and the off-diagonal terms (typically set to zero) are equivalent to dampers connecting the respective DoF of the measurement locations. To retain a physical interpretation, the constraints on $\mathbf{E}$ are symmetry and positive definiteness. Also, the corrective forces $\mathbf{c}_{2}^{T}\mathbf{E}\dot{y}(t)$ are proportional to the velocity measurements and added grounded dampers. The velocity measurements $\dot{y}(t)$ can be obtained by integration of acceleration measurements $\ddot{y}(t)$ in Equation \ref{ACC}. The integration might add long period drifts in velocity measurements, and high-pass filtering can be performed to remove these baseline shifts.} To determine $\textbf{E}$, the objective function to be minimized is the trace of the estimation error covariance matrix. Since for a general nonlinear multi-variable case, a closed-form solution for the optimal matrix $\mathbf{E}$ has not been found, a numerical optimization algorithm is used. To derive the optimization objective function, Equation \ref{NMBO} is linearized as follows
		\begin{eqnarray}\label{LNMBO}
		\mathbf{M}\ddot{\hat{q}}(t)+(\mathbf{C}_{\xi}+\mathbf{c}_{2}^{T}\mathbf{E}\mathbf{c}_{2})\dot{\hat{q}}(t)+\mathbf{K}_0{\hat{q}}(t)=\mathbf{c}_{2}^{T}\mathbf{E}\dot{y}(t)
		\end{eqnarray}
		where  \color{black}$\mathbf{K}_0$ is the initial stiffness matrix}. By defining the state error as $e=q-\hat{q}$, {\color{black}it was shown in \cite{hernandez2011natural} that} the PSD of estimation error, $\pmb{\Phi}_{ee}$, is given by
	\begin{eqnarray}\label{Phie}
	\pmb{\Phi}_{ee}(\omega)=\mathbf{H}_o\mathbf{b}_2\pmb{\Phi}_{ww}(\omega)\mathbf{b}_2^{T}\mathbf{H}_o^*+\mathbf{H}_o\mathbf{c}_2^T\mathbf{E}\pmb{\Phi}_{vv}(\omega)\mathbf{E}^T\mathbf{c}_2\mathbf{H}_o^*
	\end{eqnarray}
	with $\mathbf{H}_o$ defined as
	\begin{eqnarray}
	\mathbf{H}_o=\left(-\mathbf{M}\omega^2+\left(\mathbf{C}_{\xi}+\mathbf{c}_2^T\mathbf{Ec}_2\right)i\omega+\mathbf{K}_0\right)^{-1}
	\end{eqnarray}
	where the matrices $\pmb{\Phi}_{ww}(\omega)$ and $\pmb{\Phi}_{vv}(\omega)$ are the PSDs of the uncertain excitation on the system and measurement noise, respectively. {\color{black} In this paper, the uncertain input corresponds to the ground motion excitation, and the measurement noise corresponds to unmeasured excitations and (or) modeling errors}. To select the optimal value of $\mathbf{E}$ matrix, the following optimization problem must be solved
	{\color{black}\begin{eqnarray}\label{J}
		\begin{aligned}
		&  \argminA_{\mathbf{E} \, \in \, \mathbb{R}^{+}}  \, tr(\mathbf{P})\\
		\end{aligned}
		\end{eqnarray}}
	where $\mathbf{P}$ is the covariance matrix of displacement estimation error described as
	\begin{eqnarray}
	\begin{aligned}
	\mathbf{P}&=\mathbb{E}\left[[q(t)-\hat{q}(t)][q(t)-\hat{q}(t)]^T\right]=\int_{-\infty}^{+\infty}\pmb{\Phi}_{ee}(\omega)d\omega
	\end{aligned}
	\end{eqnarray}
	{\color{black}One alternative for the optimization problem in Equation \ref{J} can be defined if the objective is minimization of the inter-story drifts (ISD) estimation error, $\mathbf{P_\text{ISD}}$, given by
		\begin{eqnarray}\label{JT}
		\begin{aligned}
		& \argminA_{\mathbf{E} \, \in \, \mathbb{R}^{+}}  \, tr(\mathbf{P_\text{ISD}}) 
		\end{aligned}
		\end{eqnarray}}
	where 
	{\color{black}\begin{eqnarray}
		\begin{aligned}
		tr(\mathbf{P_\text{ISD}}) = \sum_{k=1}^n\mathbf{P_\text{ISD}}_{(k,k)} = \mathbf{P}_{(1,1)}+\sum_{k=2}^n[\mathbf{P}_{(k,k)}+\mathbf{P}_{(k-1,k-1)}-2\mathbf{P}_{(k,k-1)}]
		\end{aligned}
		\end{eqnarray}}
	$k$ is story number and $n$ is total number of stories.
	
\begin{figure}[!ht]
		\centering
		\includegraphics[width=0.75\linewidth]{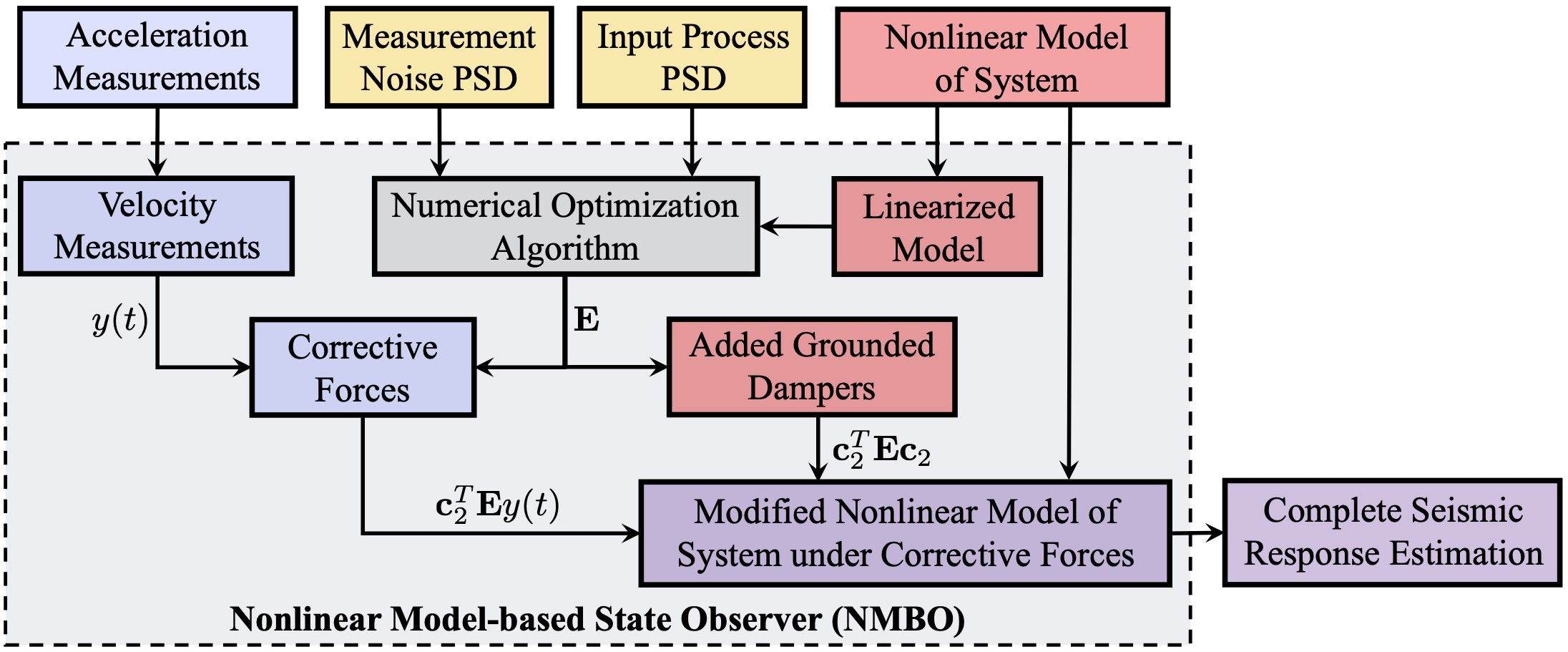}
		\caption{Summary of the nonlinear model-data fusion using the NMBO}
		\label{fig:NMBO}
	\end{figure}

\begin{figure}[!ht]
		\centering
		\includegraphics[width=0.8\linewidth]{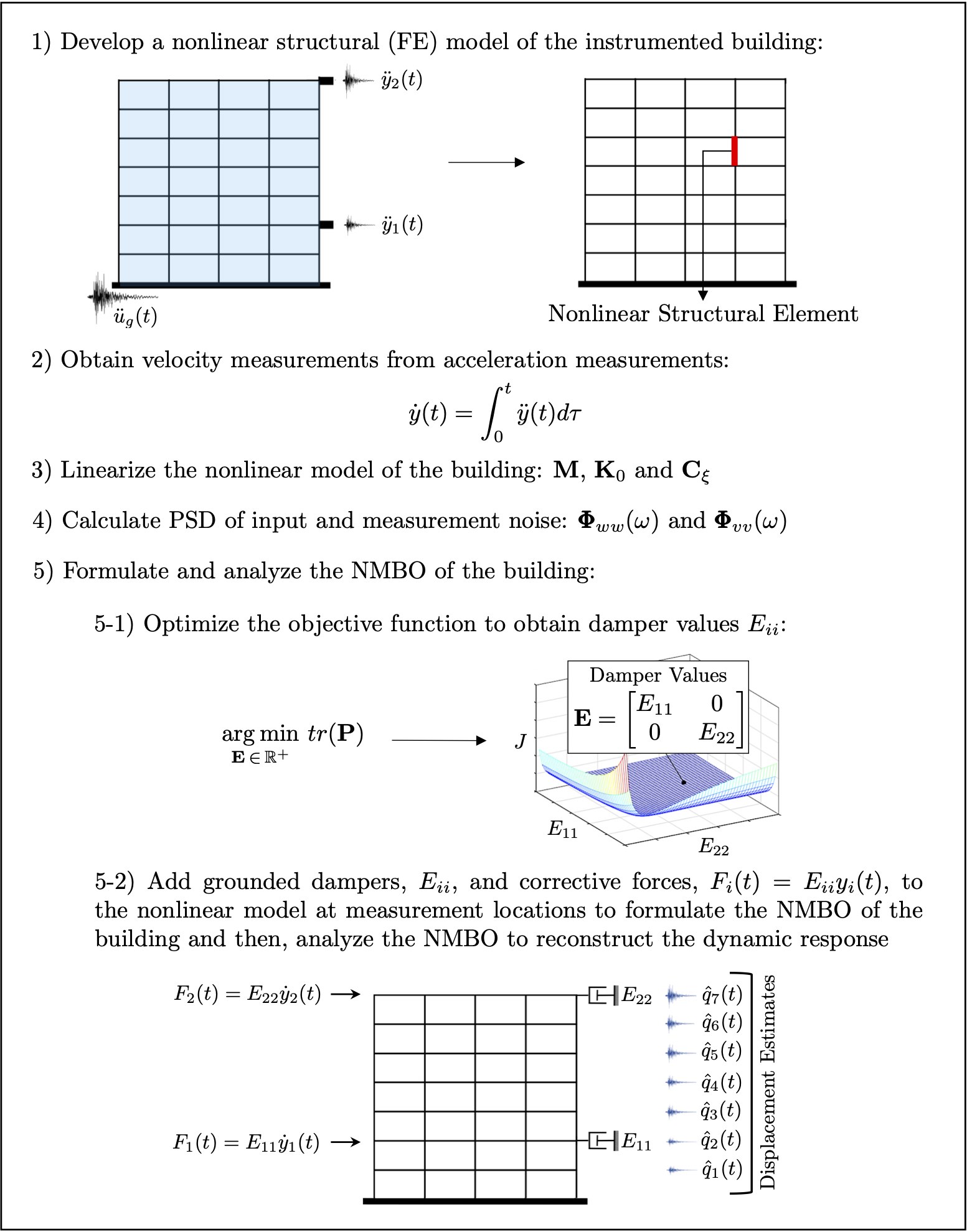}
		\caption{Implementation of the proposed nonlinear model-based observer}
		\label{fig:EMBOflowchart}
	\end{figure}

	{\color{black}Any optimization algorithm (e.g., Matlab \say{\textit{fminsearch}}) can be used to solve the optimization in Equations \ref{J} and \ref{JT} by varying the values of the diagonal elements of the $\mathbf{E}$ matrix to determine the optimized feedback matrix. {\color{black} Figure \ref{fig:NMBO} presents a summary of the nonlinear model-data fusion using the NMBO and Figure \ref{fig:EMBOflowchart} schematically illustrates the implementation of the NMBO.} Also, readers are kindly referred to \cite{hernandez2011natural,hernandez2013optimal,hernandez2018estimation,Roohi2019nonlinear,roohi2019performance}
		for implementation examples.} 
	
	\section{Proposed Seismic Monitoring Framework}
\begin{figure}[!ht]
		\centering
		\includegraphics[width=0.9\linewidth]{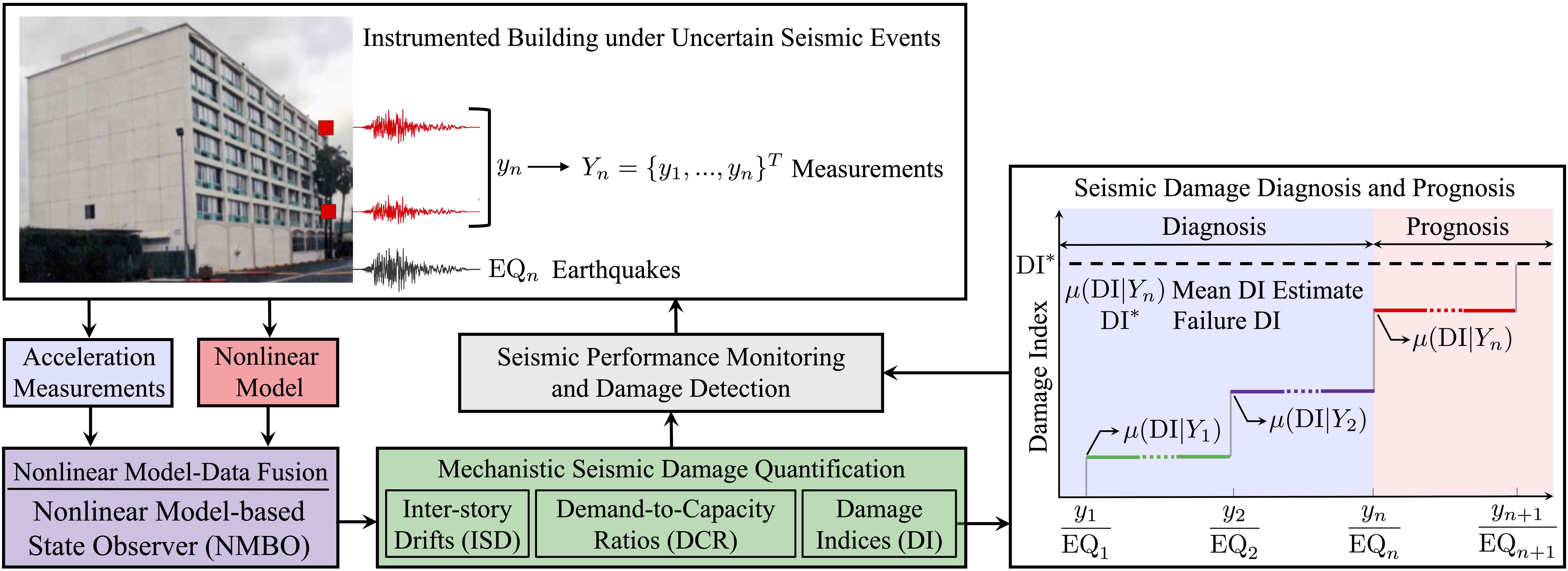}
		\caption{Summary of the proposed mechanistic damage quantification and seismic monitoring framework}
		\label{fig:fm11}
	\end{figure}

	This paper proposes a seismic monitoring framework that can be accurately employed for seismic damage detection and localization in instrumented buildings subjected to seismic ground motions. This framework employs the NMBO to combine a nonlinear structural model with acceleration measurements for reconstructing the complete seismic response. Then, the estimated response is processed to 1) estimate inter-story drifts and determine the post-earthquake re-occupancy classification of the building based on performance-based criteria 2) to compare the estimated demands with code-based capacity and reconstruct element-by-element demand-to-capacity ratios and 3) reconstruct element-level dissipated energy and ductility. The outcome of this process is employed for the performance-based monitoring, damage detection, and localization in instrumented buildings.  Figure \ref{fig:fm11} depicts a summary of the proposed seismic monitoring framework. The following subsections discuss each step of the framework in more detail.
	
	\subsubsection{Performance-based assessment using Inter-story Drifts}\label{Section:ISD}
	The {\color{black}maximum} inter-story drift ({\color{black}$\text{ISD}_{\mathrm{max}}$}) estimate at each story can be calculated using the NMBO displacement estimates as follows      \DeclarePairedDelimiter\abs{\lvert}{\rvert}
	\newcommand{\Mypm}{\mathbin{\tikz [x=1.4ex,y=1.4ex,line width=.1ex] \draw (0.0,0) -- (1.0,0) (0.5,0.08) -- (0.5,0.92) (0.0,0.5) -- (1.0,0.5);}}%
	\begin{eqnarray}\label{Drift}
	\text{ISD}_{\mathrm{max}_{k}}=\frac{\mathrm{max}\abs[\Big]{\hat{q}_k(t) - \hat{q}_{k-1}(t)}}{h_k}
	\end{eqnarray}
	where $h_k$ is height of $k$-th story, and the uncertainty in ISD estimation can be calculated as follows 
	\begin{eqnarray}\label{Driftsigma}
	\text{ISD}_{\mathrm{max}_{k}}{\:\Mypm\:\sigma_{{\text{ISD}_k}}}={\mathrm{max}\abs[\Big]{\text{ISD}_{k}{\:\Mypm\: \sqrt{\mathbf{{P}}_{\text{ISD}_{k}}}}}}
	\end{eqnarray}
	where $\sigma_{{\text{ISD}_k}}$ is the uncertainty standard deviation of ISD estimation for $k$-th story.
	The estimated ISDs are used to perform the post-earthquake evaluation of the building based on \cite{FEMA-356} performance measures, including immediate occupancy (IO), life safety (LS), and collapse prevention (CP).
	
	\subsubsection{Demand-to-capacity ratio reconstruction}
	The demand-to-capacity ratio (DCR)  for $i$-th element is reconstructed as follows
	\begin{eqnarray}\label{DCRcalculation}
	\mathrm{DCR}_{i}=\frac{\mathrm{max}|\hat{S}_i(t)|}{R_{i}}
	\end{eqnarray}
	where $\hat{S}_i(t)$ and $R_{i}$ are the seismic demand and capacity estimates of any pertinent failure mode in $i$-th structural element.
	
	\subsubsection{Dissipated energy reconstruction for damage detection and localization}\label{Section:DI}
	The seismic damage index (DI) is reconstructed using a Park-Ang type damage model \cite{park1985mechanistic} expressed as
	\begin{eqnarray}\label{DIpaper}
	DI =DI_{\mu}+DI_{E}= \dfrac{\mu_m}{\mu_u}+\psi\dfrac{E_h}{E_{max}}
	\end{eqnarray}
	where $DI_{\mu}$ and $DI_{E}$ represent damage due to excessive deformation and dissipated hysteretic energy, respectively; $\mu_m$ is the maximum ductility experienced during the earthquake, $\mu_u$ is the ultimate ductility capacity under monotonic loading, $\psi$ is calibration parameter, and $E_{max}$ is the maximum hysteretic energy dissipation capacity for all relevant failure modes. 
	
	\section{Case-study: Van Nuys Hotel Testbed}
	The proposed methodology is validated in the remaining sections using seismic response measurements from Van Nuys hotel. The CSMIP instrumented this building as Station 24386, and the recorded data of this building are available from several earthquakes, including 1971 San Fernando, 1987 Whittier Narrows, 1992 Big Bear, and 1994 Northridge earthquakes. From these data, measurements during 1992 Big Bear and 1994 Northridge earthquakes are used in this study to demonstrate the proposed framework. Researchers have widely studied the Van Nuys building \cite{Islam1996,Loh1996,Li1998,Browning2000,Taghavi2005,Goel2005,Bernal2006,ching2006bayesian,Naeim2006,Todorovska2008,Rodriguez2010damage,Gicev2012,Trifunac2014detection,Shan2016model,Pioldi2017} and the building was selected as a testbed for research studies by researchers in PEER \cite{Krawinkler2005}. 
	
	\subsection{Description of the Van Nuys building}
	The case-study building is a 7-story RC building located in San Fernando Valley in California. The building plan is 18.9 m $\times$ 45.7 m in the North-South and East-West directions, respectively. The total height of the building is 19.88 m, with the first story of 4.11 m tall, while the rest are 2.64 m approximately. The structure was designed in 1965 and constructed in 1966. Its vertical load transfer system consists of RC slabs supported by concrete columns and spandrel beams at the perimeter. The lateral resisting systems are made up of interior concrete column-slab frames and exterior concrete column-spandrel beam frames. The foundation consists of friction piles, and the local soil conditions are classified as alluvium. The testbed building is described in more detail in \cite{Trifunac1999,Krawinkler2005}.
	
	\subsection{Building instrumentation}
\begin{figure}[H]
		\centering
		\includegraphics[width=0.9\linewidth]{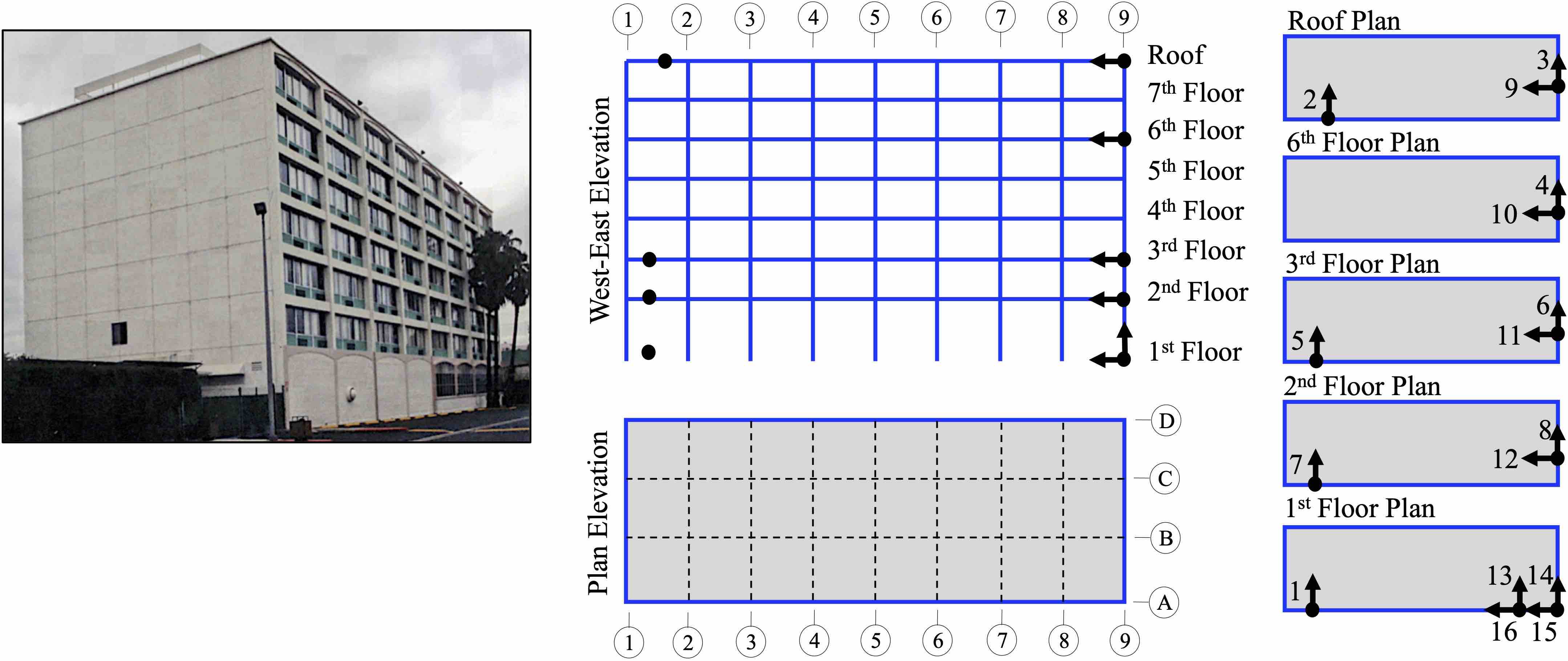}
		\vspace{-5pt}
		\caption{(left) Van Nuys hotel testbed (CSMIP Station 24386) and (Right) Location of building accelerometers on the West-East elevation and floor plans}
		\label{fig:Sensors}
	\end{figure}

	The CSMIP initially instrumented the building with nine accelerometers at the 1st, 4th, and roof floors. Following the San Fernando earthquake, CSMIP replaced the recording layout by 16 remote accelerometer channels connected to a central recording system. These channels are located at 1st, 2nd, 3rd, 6th, and roof floors. Five of these sensors measure longitudinal accelerations, ten of them measure transverse accelerations, and one of them measures the vertical acceleration. Figure \ref{fig:Sensors} shows the location of accelerometers.
	
	\subsection{Earthquake damage}
\begin{figure}[H]
		\centering
		\includegraphics[width=1.0\linewidth]{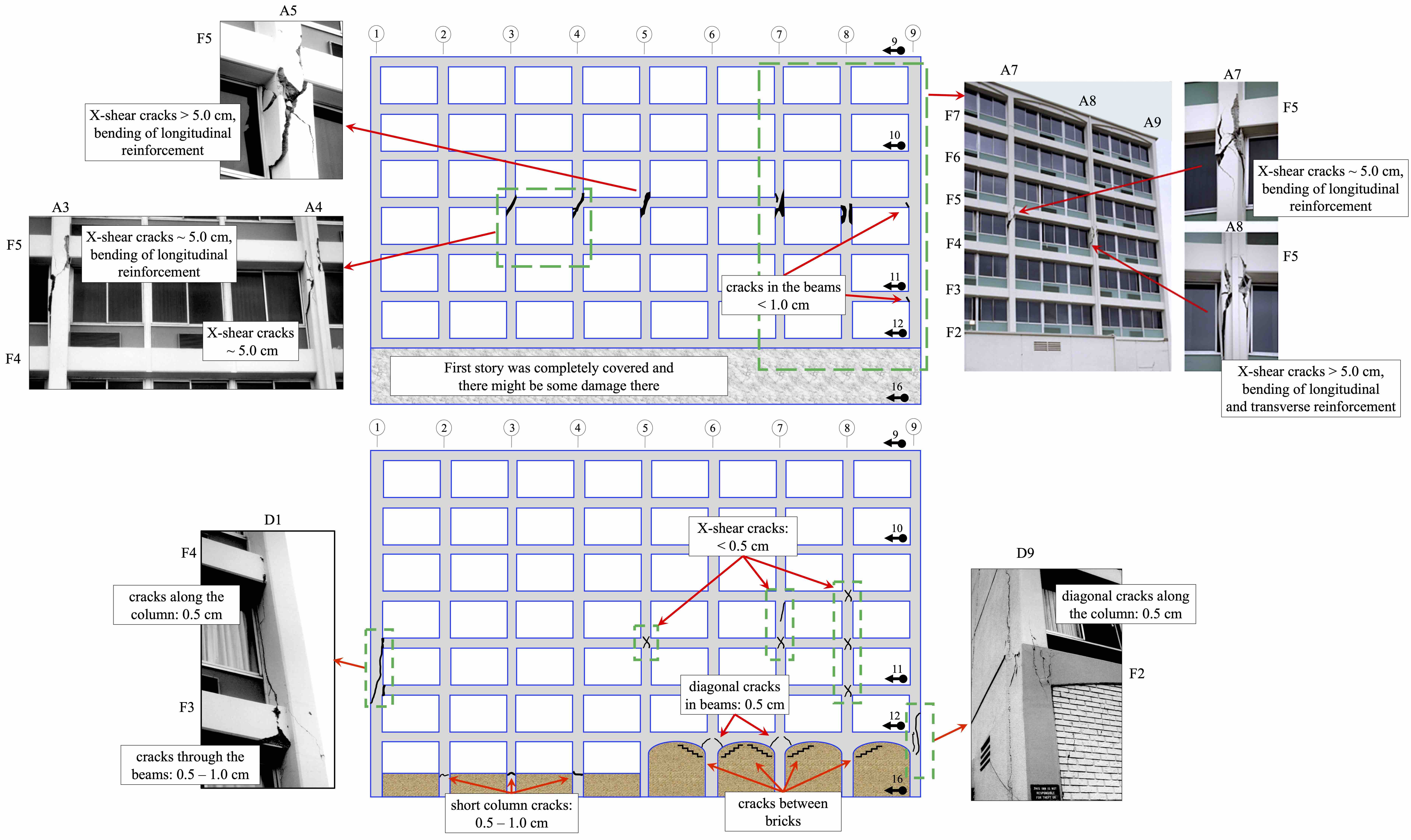}
		\vspace{-5pt}
		\caption{Schematic representation and photo records of of seismic damage following the 1994 Northridge earthquake: (top) south view of Frame A, and (bottom) south view of Frame D. (Adopted from Trifunac and Ivanovic 2003)}
		\label{fig:CrackPicks}
	\end{figure}

	Since the Van Nuys building was instrumented and inspected following earthquakes that affected the structure, the history of damage suffered by this building is well-documented. These documents show that the building has experienced insignificant structural and mostly nonstructural damage before the Northridge earthquake in 1994. However, the Northridge earthquake extensively damaged the building. Post-earthquake inspection red-tagged the building and revealed that the damage was severe in the south longitudinal frame (Frame A). In Frame A, five of the nine columns in the 4th story (between floors 4 and 5) were heavily damaged due to inadequate transverse reinforcement, and shear cracks ($\geq 5cm$) and bending of longitudinal reinforcement were easily visible \cite{Trifunac2003}.  Figure \ref{fig:CrackPicks} demonstrate the seismic damage following the 1994 Northridge earthquake in the south and north frames.
	
	\subsection{Previous damage assessment studies on Van Nuys building}
	\cite{Browning2000} reported the performance assessment results of the Van Nuys building based on studies of three independent research teams.  These teams used nonlinear dynamic and nonlinear static analysis to localize structural damage and concluded that the various studies were successful to varying degrees. \cite{Naeim2006} presented a methodology for automated post-earthquake damage assessment of instrumented buildings. The methodology was applied to the measured response from Landers, Big Bear, and Northridge earthquakes. Their findings show that the building did not suffer structural damage under the Landers and Big Bear earthquakes and indicate a high probability of extensive damage to the middle floors of the building under the Northridge earthquake. They concluded that their methodology was not able to identify the exact floor level at which the damage occurs because no sensors were installed on the floor that was damaged. As previously mentioned; \cite{ching2006bayesian} performed state estimation using measured data during the Northridge earthquake combined with a time-varying linear model and then, with a simplified time-varying nonlinear degradation model derived from a nonlinear finite-element model of the building. They found that state estimation using the nonlinear degradation model shows better performance and estimates the maximum ISD to be at the 4th story. They concluded that an appropriate estimation algorithm and a suitable identification model can improve the accuracy of the state estimation. \cite{Todorovska2008} used impulse response functions computed from the recorded seismic response during 11 earthquakes, including the Northridge earthquake. They analyzed travel times of vertically propagating waves to obtain the degree and spatial distribution of changes in stiffness and infer the presence of structural damage. Their findings showed that during the Northridge earthquake, the rigidity decreased by about 60\% between the ground and 2nd stories; by about 33\% between 2nd and 3rd stories, and between 3rd and 6th stories; and by about 41\% between the 6th story and roof. \cite{Rodriguez2010damage} implemented their proposed method called Baseline Stiffness Method to detect and assess structural and nonstructural damage to the Van Nuys building using data from the Northridge earthquake. Their approach was able to detect damage in connections with wide cracks of 5 cm or greater. On the other hand, the method identified damage in some elements of upper stories that were not detected by visual inspection reports, and also, they could not identify some of the moderate damages with small cracks. \cite{Shan2016model} presented a model-reference damage detection algorithm of hysteretic buildings and investigated the Van Nuys hotel using measured data from Big Bear and Northridge earthquakes. The researchers concluded that their algorithm can only detect damages of certain floors and cannot detect damages in structural components or connections of the instrumented structure.
	
	\section{Implementation of the Proposed Seismic Monitoring Framework}
	\subsection{Nonlinear modeling of the Van Nuys hotel testbed in OpenSEES}
	The nonlinear FE model of the building was implemented using a two-dimensional fixed-base model within the environment of OpenSEES \cite{opensees}. This model corresponds to one of the longitudinal frames of the building (Frame A in Figure \ref{fig:Sensors}). In the FE model, beams and columns were modeled based on distributed plasticity modeling approach, and the \textit{force-based beam-column} elements were used to accurately determine yielding and plastic deformations at the integration points along the element. Gauss-Lobatto integration approach was employed to evaluate the nonlinear response of force-based elements. Each beam and column element was discretized with four integration points, and the cross-section of each element was subdivided into fibers. The uniaxial \textit{Concrete01} material was selected to construct a \textit{Kent-Scott-Park} object with a degraded linear unloading and reloading stiffness and zero tensile strength. The uniaxial \textit{Steel01} material was used to model longitudinal reinforcing steel as a bilinear model with kinematic hardening. The elasticity modulus and strain hardening parameters were assumed to be 200 GPa and 0.01, respectively. Due to insufficient transverse reinforcement in beams and columns \cite{Jalayer2017}, an unconfined concrete model was defined to model concrete. The peak and post-peak strengths were defined at a strain of 0.002 and a compressive strain of 0.006, respectively. The corresponding strength at ultimate strain was defined as $0.05f'_c$ for $f'_c= 34.5$ MPa and $f'_c= 27.6$ MPa and $0.2f'_c$ for $f'_c= 20.7$ MPa. Based on the recommendation of \cite{Islam1996}, the expected yield strength of Grade 40 and Grade 60 steel were defined as 345 MPa (50 ksi) and 496 MPa (72 ksi), respectively, to account for inherent overstrength in the original material and strength gained over time.
	
	\subsection{Formulation of the OpenSEES-NMBO of Van Nuys building}
	The nonlinear FE model and response measurements of the Van Nuys building was employed to implement the NMBO in OpenSEES. The following subsections present the step-by-step formulation of the OpenSEES-NMBO.
	
	\subsubsection{PSD selection and numerical optimization}
	{\color{black}The PSD of ground motion, $\pmb{\Phi}_{ww}(\omega)$, was characterized using the Kanai-Tajimi PSD given by
		\begin{equation}\label{KT}
		S(\omega)=G_0\frac{1+4\xi_g^{2}(\frac{\omega}{\omega_g})^{2}}{\left[1-(\frac{\omega}{\omega_g})^2\right]^2+4\xi_g^{2}(\frac{\omega}{\omega_g})^{2}}
		\end{equation} 
		and the amplitude modulating function $I(t)$ was selected as
		\begin{equation}\label{It}
		I(t)=te^{-\alpha{t}}
		\end{equation}}
	The parameter were defined as $\xi_g=0.35$ for both earthquakes, $\omega_g=6\pi rad/s$ for Northridge earthquake and $\omega_g=2\pi rad/s$ for Big Bear earthquake. The underlying white noise spectral density $G_0$ for each direction of measured ground motion for each shake table test was found such that about 95\% of the Fourier transform of the measured ground motion lies within two standard deviations of the average from the Fourier transforms of an ensemble of 200 realizations of the Kanai-Tajimi stochastic process. $\alpha$ was selected as 0.12. Details can be found in \cite{Roohi2019nonlinear}. Also, the PSD of measurement noise, $\pmb{\Phi}_{vv}(\omega)$, in each measured channel was taken as zero mean white Gaussian sequences with a noise-to-signal root-mean-square (RMS) ratio of 0.02.
	
	Numerical optimization was performed using Equation \ref{JT}. Table \ref{table:assembly} presents the optimized damper values for each seismic event.
	
	\begin{table}
		\caption{Optimized damper values in kN.s/m (kips.s/in) units}
		\label{table:assembly}
		\centering
		\small
		\renewcommand{\arraystretch}{1.25}
		\begin{tabular}{l l l l l  }
			\hline \hline
			Earthquake & \multicolumn{1}{c}{Story 1} & \multicolumn{1}{c}{Story 2} & \multicolumn{1}{c}{Story 5} & \multicolumn{1}{c}{Story 7} \\
			\hline
			Big Bear & 7283.11 (41.59) & 9357.25 (53.43) & 19299.40 (110.20) & 34808.04 (198.76) \\
			Northridge & 5209.72 (29.75) & 6592.45 (37.64) & 16612.79 (94.86) & 47217.69 (269.62) \\
			\hline \hline
		\end{tabular}
		\normalsize
	\end{table}
	
	\subsubsection{Formulation of the OpenSEES-NMBO}
\begin{figure}
		\centering
		\includegraphics[width=1\linewidth]{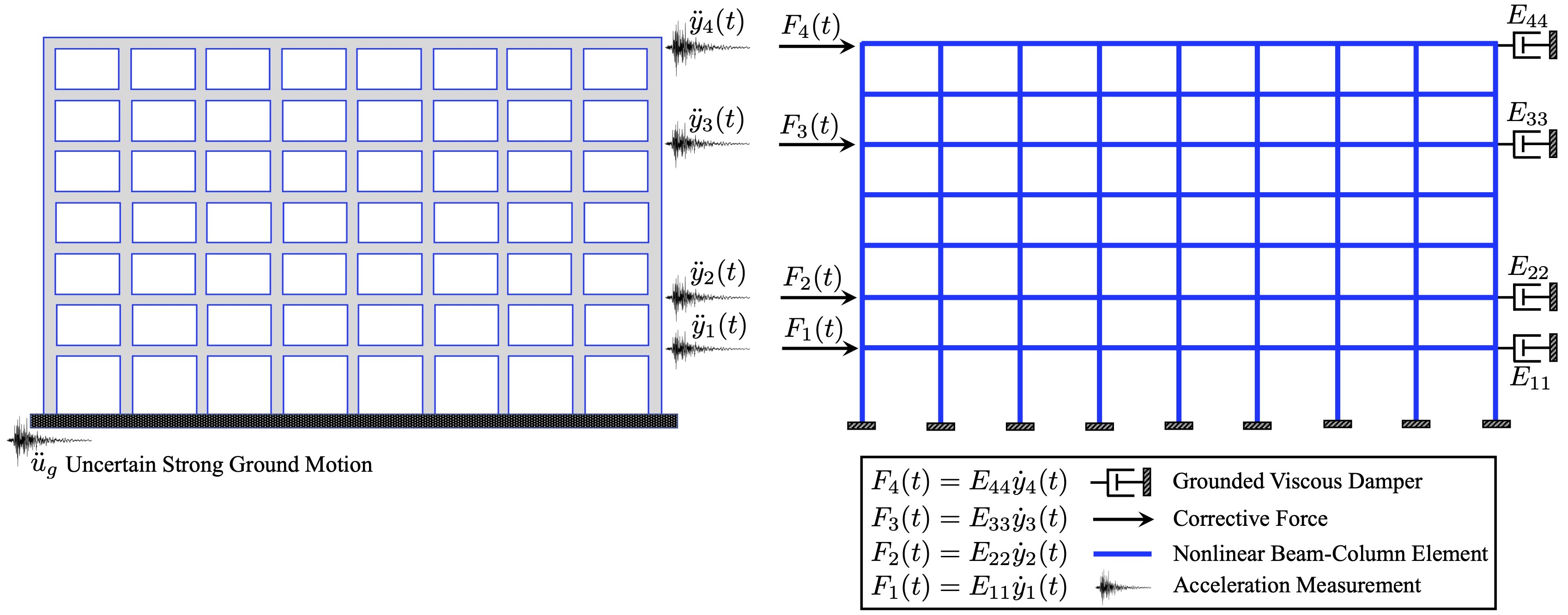}
		\caption{Schematic of the Van Nuys hotel testbed with location of accelerometers (left) and the OpenSEES-NMBO with corresponding added viscous dampers and corrective forces in measurement locations}
		\label{fig:vannuysnmbo}
	\end{figure}

	The OpenSEES nonlinear FE model was modified by adding grounded dampers in measurement locations and was subjected to corrective forces. Dynamic analysis was performed to estimate the complete seismic response. Figure \ref{fig:vannuysnmbo} presents a schematic of the Van Nuys hotel testbed (with the location of accelerometers) along with the OpenSEES-NMBO (with corresponding added viscous dampers and corrective forces in measurement locations).
	
	\subsection{Seismic damage reconstruction using estimated seismic response}\label{DQ}
	Once the complete seismic response is estimated using the OpenSEES-NMBO, the seismic damage to the building can be quantified according to the Section \ref{Section:DI}. This subsection demonstrates the procedure in more detail.
	
	\subsubsection{Shear DCR reconstruction}
	The shear DCRs were estimated based on the Equation \ref{DCRcalculation}. The shear demands were obtained from OpenSEES-NMBO, and the capacity of columns were calculated based on the Section 6.5.2.3.1 of \cite{FEMA-356}.
	
	\subsubsection{Ductility demand reconstruction}
	The seismic damage caused by excessive deformation is the first term of the Equation \ref{DIpaper} given by:
	\begin{eqnarray}
	DI_{\mu} = \dfrac{\mu_m}{\mu_u}
	\end{eqnarray}    
	where, the $\mu_m$ in each structural element is expressed by the maximum estimated curvature along integration points normalized by the yield curvature given by
	\begin{eqnarray}\label{E1}
	\mu_m = max\{ {\dfrac{\phi_{m,j}}{\phi_y}}\}_{j=1:N_p}
	\end{eqnarray}
	here, $\phi_{m,j}$ is maximum estimated curvature in integration point $j$, $\phi_y$ is curvature capacity and $N_p$ is number of integration points along element. The curvature ductility capacity ($\mu_{u}$) is obtained by
	\begin{eqnarray}\label{E2}
	\begin{aligned}
	&\mu_{u} =  \dfrac{\phi_{u}}{\phi_y}
	\end{aligned}
	\end{eqnarray}
	where $\phi_{u}$ is the ultimate curvature capacity of the section.
	
	\subsubsection{Dissipated hysteretic energy reconstruction}
	The seismic damage caused by dissipated hysteretic energy, $DI_{E}$ in Equation \ref{DIpaper}, was calculated based on flexure failure mode as follows
	\begin{eqnarray}
	DI_{E} = \psi\dfrac{E_h}{E_{max}}\cong \psi\dfrac{E_h}{M_y\theta_y\mu_u}
	\end{eqnarray}
	where $M_y$ is yield moment and $\theta_y$ is yield rotation angle. The main issue with the calculation of $DI_E$ is the determination of $\psi$, which usually is calibrated to a number between 0.05 or 0.15. A reasonable $\psi$ value should properly account for the effect of load cycles causing structural damage. The selection of small value for $\psi$ neglects the effect of the $DI_E$ in the overall damage index \cite{Williams1995}. Since the true $\psi$ is unknown for the elements of Van Nuys building and in the scope of this paper, the objective is to localize seismic damage, the calibration parameter is set to one, and the estimated value of each term in the damage index will be first reported separately and then combined.
	
	The dissipated hysteretic energy ($E_h$) is estimated based on Equation \ref{energy2} and the seismic response estimated using OpenSEES-NMBO. The parameter $M_y$ was obtained based on section analysis and the value of $\theta_y\mu_u$ calculated as follows
	\begin{eqnarray}\label{E3}
	\theta_y\mu_u = \theta_p=(\phi_u-\phi_y)l_p=\phi_pl_p
	\end{eqnarray}
	where $l_p$ is the  plastic hinge length and is defined using an  empirically  validated  relationship proposed by \cite{Bae2008} given by
	\begin{eqnarray}\label{E4}
	\dfrac{l_p}{h}=\left[0.3\left(\dfrac{P}{P_o}\right)+\left(\dfrac{A_s}{A_g}\right)-1\right]\left(\dfrac{L}{h}\right)+0.25\geq0.25
	\end{eqnarray}
	where $h$ and $L$ represent depth and length of column; $A_g$ and $A_s$ denote gross area of concrete section and area of tension reinforcement; $f'_c$ and $f_y$ are compressive strength of concrete and yield stress of reinforcement; and $P_o = 0.85f'_c (A_g-A_s) + f_y A_s$.
	
	\section{Seismic Response and Damage Reconstruction Results}
	{\color{black}A summary of the seismic response and damage reconstruction results is presented in this section to validate the proposed seismic monitoring framework.}
	\subsection{Displacement estimation results}
\begin{figure}[H]
		\centering
		\includegraphics[width=1\textwidth]{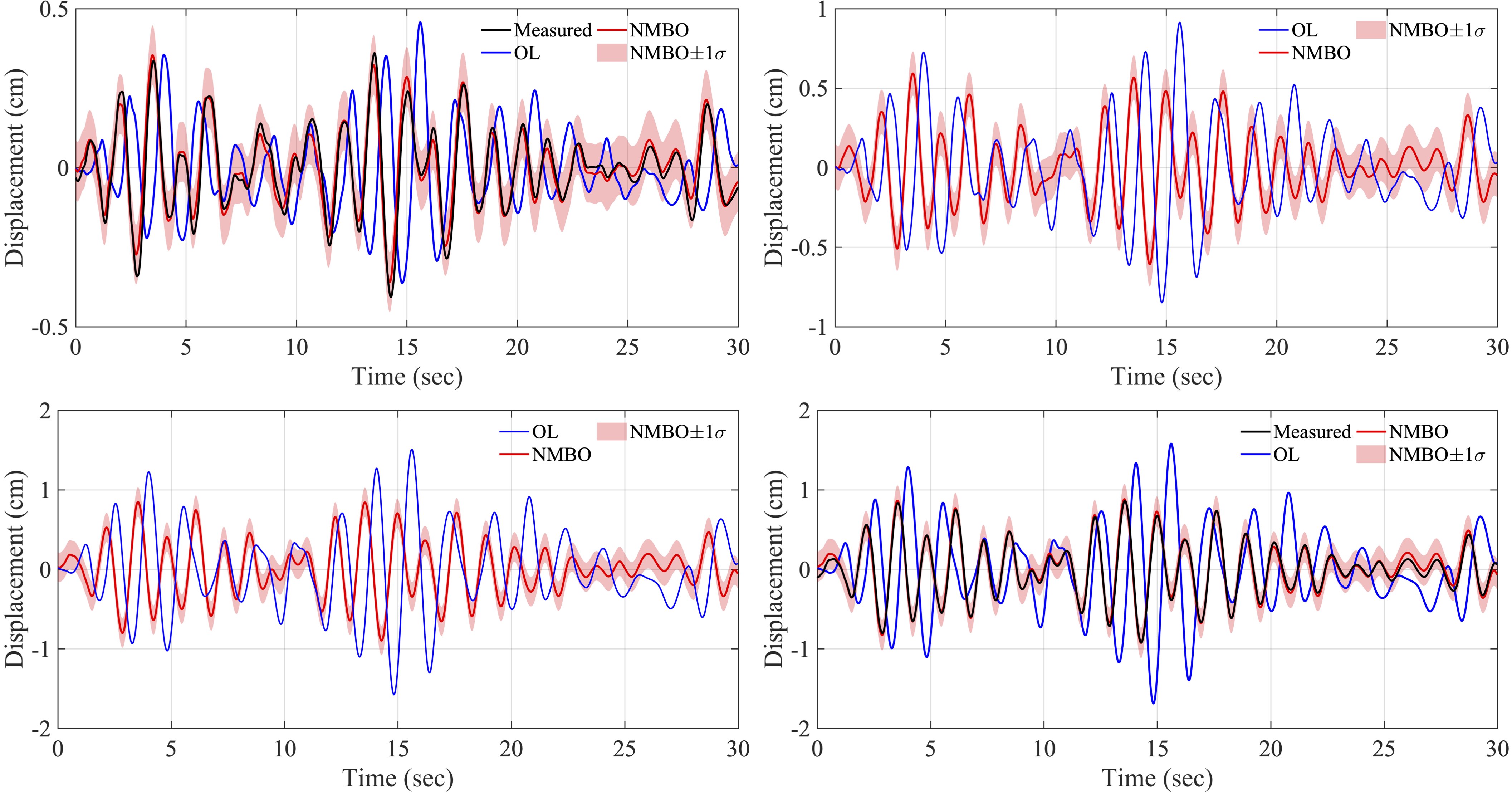}
		\caption{Comparison of displacement estimates using OpenSEES-NMBO with estimates using open-loop analysis and actual measurements in 1st floor (top left), 3rd floor (top right), 6th floor (bottom left) and 7th floor (bottom right) during Big Bear earthquake. The \textit{Measured} represents measured response, the \textit{OL} represents the open-loop analysis of OpenSEES model under measured ground motion and the \textit{NMBO} represents the estimated response using the OpenSEES-NMBO with sensor measurements from measured location along with $1\sigma$ estimation uncertainty bound}
		\label{DispBigBear}
	\end{figure}

\begin{figure}[H]
		\centering
		\includegraphics[width=1\textwidth]{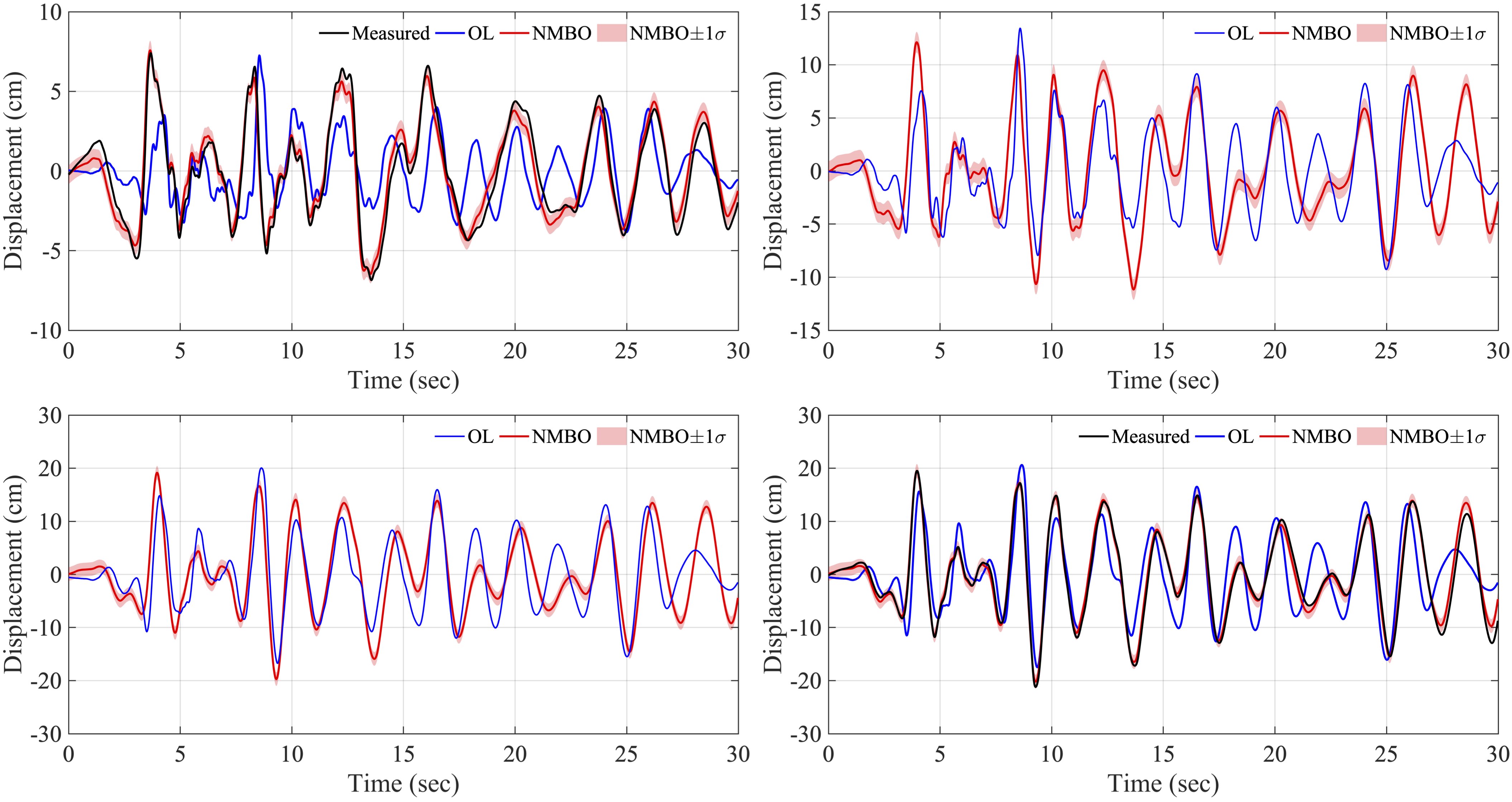}
		\caption{Comparison of displacement time history estimates with estimates using open-loop analysis and actual measurements in 1st floor (top left), 3rd floor (top right), 6th floor (bottom left) and 7th floor (bottom right) during Northridge earthquake.}
		\label{DispNorthridge}
	\end{figure}

	First, we compare the displacement estimates using OpenSEES-NMBO and its uncertainty with those obtained from 1) response measurements and 2) open-loop analysis under measured ground motion at instrumented and non-instrumented stories. Figures \ref{DispBigBear} and \ref{DispNorthridge} present the comparison of the displacement estimates at instrumented 1st and 7th stories and non-instrumented 3rd and 6th stories during the Big Bear earthquake and Northridge earthquake, respectively.
\subsection{Inter-story drift estimation results}
\begin{figure}[H]
		\centering
		\includegraphics[width=0.85\textwidth]{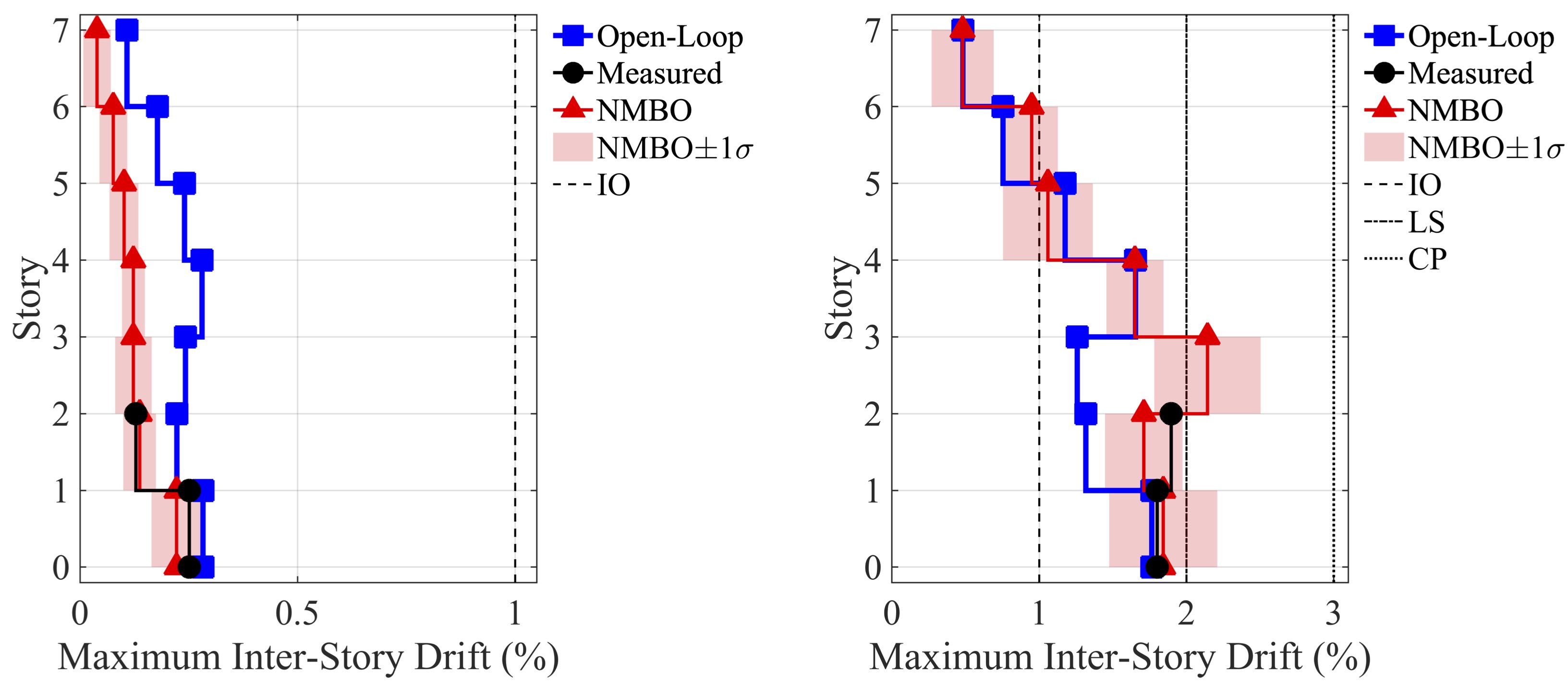}
		\caption{Comparison of $\text{ISD}_{\mathrm{max}}$ ratios obtained from response measurements with those estimated using OpenSEES-NMBO and open-loop analysis during 1992 Big Bear earthquake (left) and 1994 Northridge earthquake (right).}
		\label{Drifts}
	\end{figure}

\begin{figure}[H]
		\centering
		\includegraphics[width=1\textwidth]{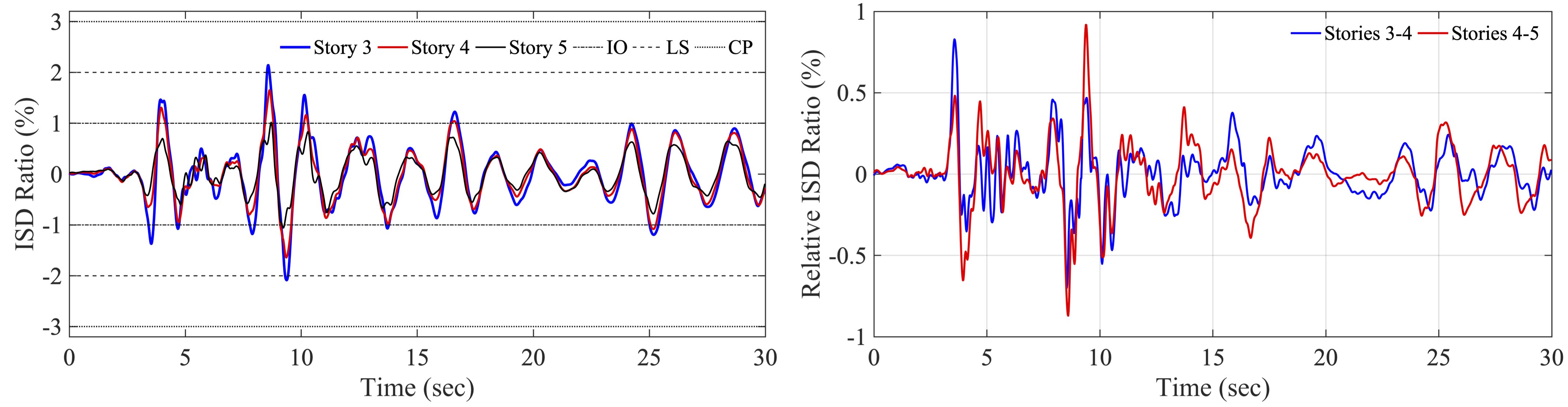}
		\caption{Comparison of ISD (left) and RISD (right) time history estimates for stories 3 to 5 during Northridge earthquake.}
		\label{RDriftNorthridge}
	\end{figure}

Figure \ref{Drifts} depicts the estimated $\text{ISD}_{\mathrm{max}}$ ratios and their corresponding $1\sigma$ confidence intervals using OpenSEES-NMBO. These results are compared with estimated $\text{ISD}_{\mathrm{max}}$ using open-loop analysis and those obtained from instrumented stories. The OpenSEES-NMBO ISD estimates indicate that the building could be classified as IO following the Big Bear earthquake and as LS-CP following the Northridge earthquake. The actual performance and post-earthquake inspection reports of the buildings validate the accuracy of the performance estimates. Figure \ref{RDriftNorthridge} gives an in-depth examination of the ISD estimates during the Northridge earthquake. The left plot in this figure shows the comparison of ISDs at 3rd, 4th, and 5th stories, and the right plot shows the comparison of relative ISDs between floors 3 and 4 and also, floors 4 and 5. Here, the relative ISD is defined as follows
\begin{eqnarray}\label{RISD}
\text{RISD}_{(k,k-1)} = \text{ISD}_{(k)}-\text{ISD}_{(k-1)}
\end{eqnarray}
where $\text{RISD}_{(k,k-1)}$ is relative ISD between stories $k$ and $k-1$. The estimation results show that even though the $\text{ISD}_{\mathrm{max}}$ occurs in the third story, the RISD demand between floors 4 and 5 is higher than floors 3 and 4.
	
	
	\subsection{Elemet-by-element shear demand to capacity ratio reconstruction}
\begin{figure}[H]
		\centering
		\includegraphics[width=1\textwidth]{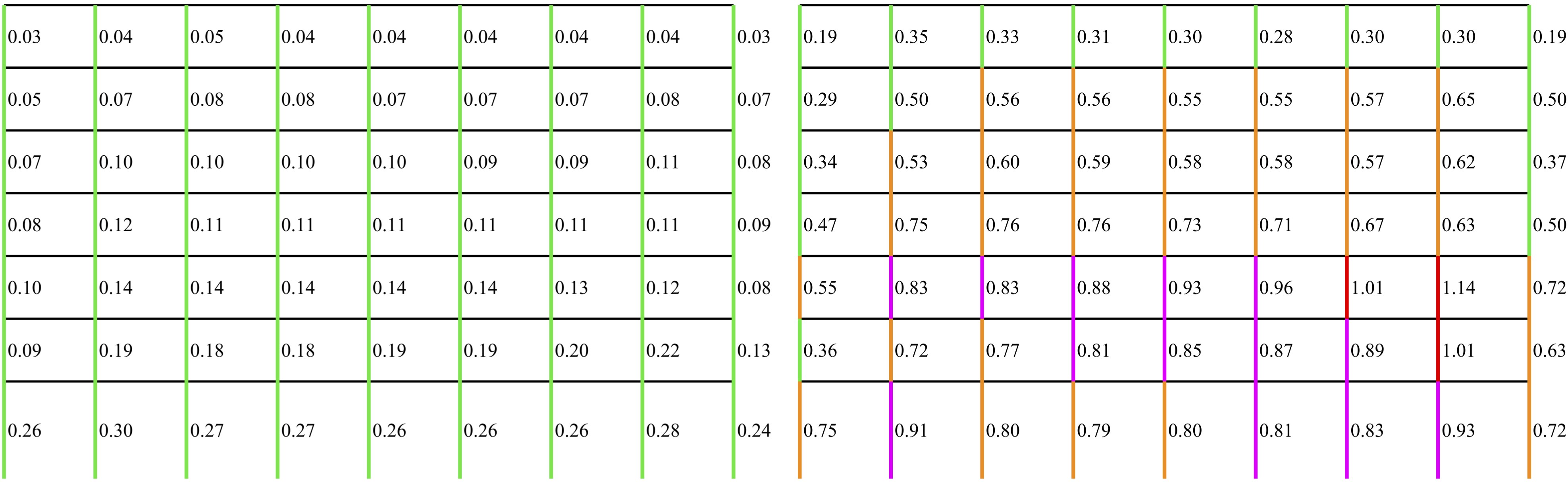}
		\caption{Estimated element-by-element shear demand to capacity ratios by OpenSEES-NMBO using measured seismic response of the Van Nuys building during 1992 Big Bear (left) and 1994 Northridge (right) earthquakes.}
		\label{Shear}
	\end{figure}

	Figure \ref{Shear} shows results for shear estimated element-by-element shear DCR ratios by OpenSEES-NMBO using measured seismic response of the Van Nuys building during Big Bear (left) and Northridge (right) earthquakes.
	
	\subsection{Element-by-element damage index reconstruction}
\begin{figure}[H]
		\centering
		\includegraphics[width=1\textwidth]{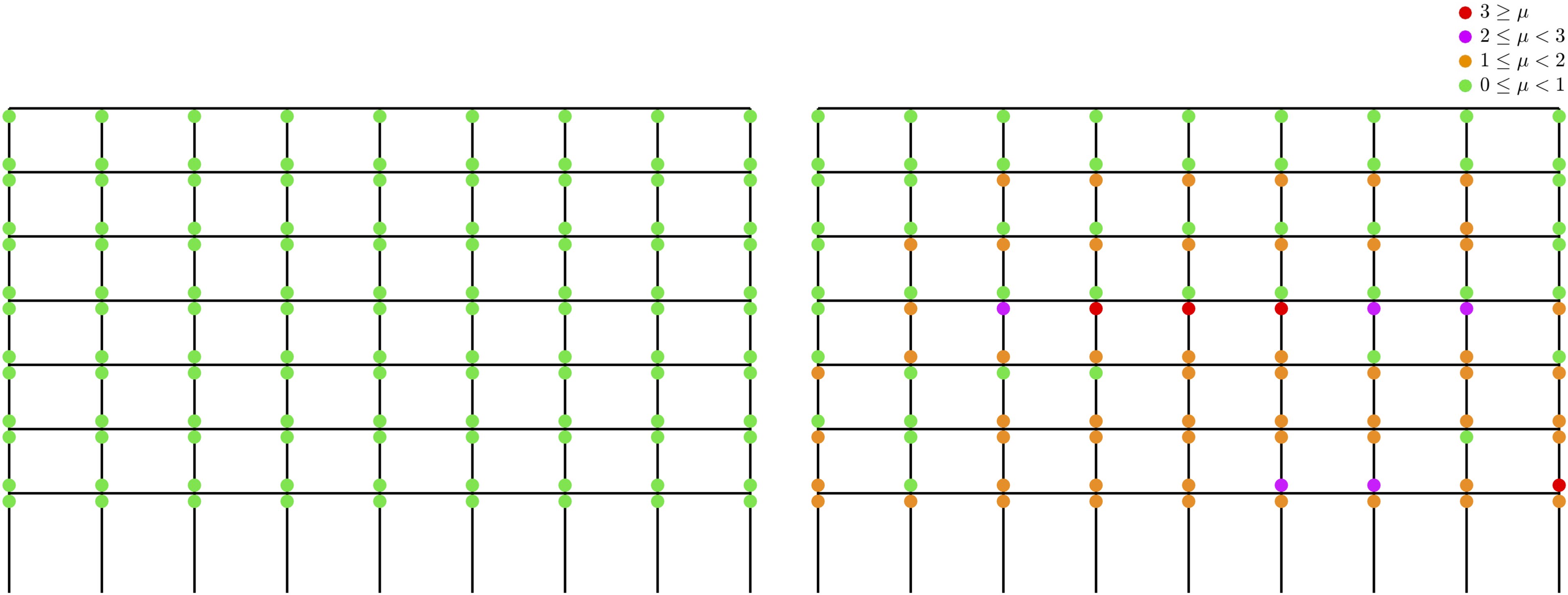}
		\caption{Reconstructed maximum end curvature ductility demands in columns by implementing OpenSEES-NMBO using measured seismic response of the Van Nuys building during 1992 Big Bear (left) and 1994 Northridge (right) earthquakes.}
		\label{curvature}
	\end{figure}

\begin{figure}[H]
		\centering
		\includegraphics[width=1\textwidth]{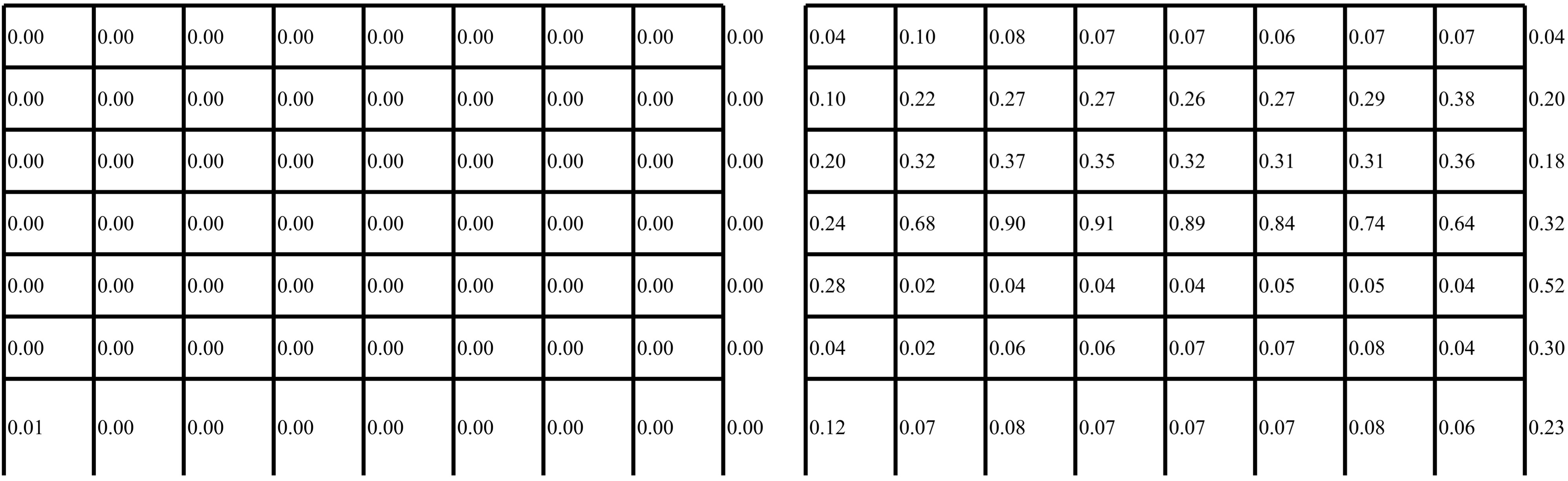}
		\caption{Reconstructed element-by-element normalized energy dissipation by OpenSEES-NMBO using measured seismic response of the Van Nuys building during 1992 Big Bear (left) and 1994 Northridge (right) earthquakes.}
		\label{energyNMBO}
	\end{figure}

\begin{figure}[H]
		\centering
		\includegraphics[width=1\textwidth]{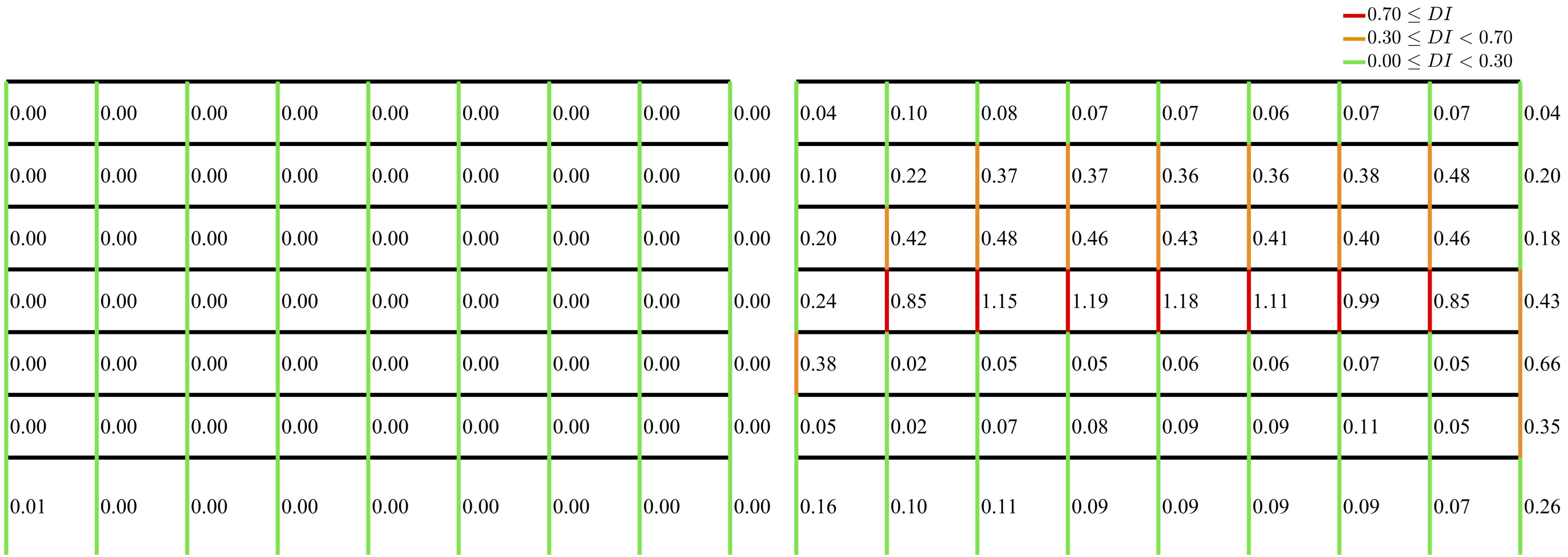}
		\caption{Reconstructed element-by-element damage indices by OpenSEES-NMBO using measured seismic response of the Van Nuys building during 1992 Big Bear (left) and 1994 Northridge (right) earthquakes.}
		\label{DI}
	\end{figure}

\begin{figure}[H]
		\centering
		\includegraphics[width=1\linewidth]{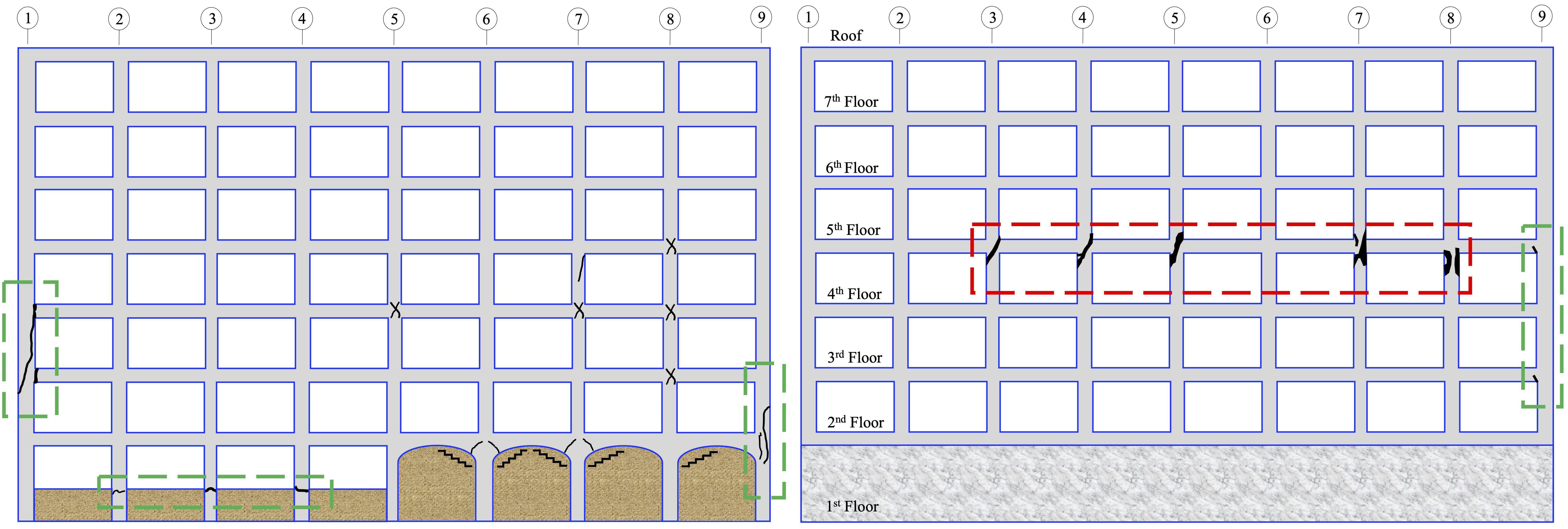}
		\vspace{-5pt}
		\caption{Seismic damage experienced during the 1994 Northridge earthquake: (left) south view of Frame D, and (right) south view of Frame A. (Adopted from Trifunac and Ivanovic 2003)}
		\label{fig:Cracks1}
	\end{figure}

	This section presents the seismic damage quantification results using the estimated response from OpenSEES-NMBO and the damage model presented at Section \ref{Section:DI}, which is also demonstrated in more detail in Section \ref{DQ}. Figure \ref{curvature} summarizes the estimated maximum curvature ductility demands ($\mu_m$) in two ends of columns for each earthquake. To interpret the $\mu_m$ demands, one needs to consider that the expected ductility capacity of columns in this building is relatively low as the columns are non-ductile. Figure \ref{energyNMBO} presents reconstructed element-by-element normalized energy dissipation. Figure \ref{DI} presents the reconstructed element-by-element damage indices. Figure \ref{fig:Cracks1} schematically depicts the seismic damage suffered during the Northridge earthquake to compare the reconstructed DIs with the building's actual performance. The shear cracks with $\text{width} \geq 5cm$ are highlights in red color and the shear cracks ($0.5cm \leq \text{width}\leq 1$) are highlights in green color. As can be seen, the element-by-element comparison of estimated DIs with post-earthquake inspection results confirms the accuracy of damage localization using the proposed mechanistic approach.
	
	\subsection{Discussion on damage detection and localization results}
	The results described in the preceding sections demonstrate that a nonlinear model-data fusion using a refined distributed plasticity FE model and a limited number of response measurements can accurately reconstruct the seismic response. Subsequently, the estimated response can be used to quantify the seismic damage based on damage sensitive response parameters and damage models. The estimated ISDs indicated that the performance-based post-earthquake re-occupancy category of the building was IO during the Big Bear earthquake and LS-CP during the Northridge earthquake. The ISD and RISD analysis during the Northridge earthquake showed that the $\text{ISD}_{\mathrm{max}}$ occurred at the 3rd story, while the maximum RISD occurs at the top of the 4th story. Also, dissipated energy and ductility reconstruction detects no structural damage during the Big Bear earthquake and severe damage during the Northridge earthquake. By combining the information from estimated ISDs, RISDs, maximum curvature ductility demands, and element-by-element damage indices during the Northridge earthquake, severe damage was localized in the columns of the 4th story (between floors 4 and 5) and also, small or moderate damage was estimated for the remaining columns. The location of severe damage in the 4th story can be explained mainly by widely spaced or absent transverse reinforcing in the beam-column joints contributed to the lower shear capacity of the story; which can be accounted for by the proposed mechanistic seismic monitoring framework through high-resolution seismic response and element-by-element damage index reconstruction. Finally, it was shown that the damage assessment results were consistent with the building's actual performance and post-earthquake inspection reports following the Big Bear and Northridge earthquakes. Therefore, the applicability of the proposed framework is validated in the context of a real-world building that experienced severe localized damage during sequential seismic events.
	
	\section{Conclusions}
	This paper proposes a seismic monitoring framework to reconstruct element-by-element dissipated hysteretic energy and perform structural damage detection and localization. The framework employs a  nonlinear model-based state observer (NMBO) to combine a design level nonlinear FE model with acceleration measurements at limited stories to estimate nonlinear seismic response at all DoF of the model. The estimated response is then used to reconstruct damage-sensitive response features, including 1) inter-story drifts, 2) code-based demand to capacity ratios, and 3) normalized dissipated hysteretic energy and ductility demands. Ultimately, the estimated features are used to conduct the performance-based post-earthquake assessment, damage detection, and localization. 
	
	The methodology was successfully validated using measured data from the seven-story Van Nuys hotel testbed instrumented by CSMIP (Station 24386) during 1992 Big Bear and 1994 Northridge earthquakes. The NMBO of the building was implemented using a distributed plasticity finite element model and measured data to reconstruct seismic response during each earthquake. The estimated seismic response was then used to reconstruct inter-story drifts and determine the performance-based post-earthquake re-occupation category of the building following each earthquake. The performance categories were estimated as IO and LS-CP during the Big Bear and Northridge earthquakes, respectively. Analysis during the Northridge earthquake showed that the maximum inter-story drift occurred at the 3rd story, while the maximum relative inter-story drift occurred at the top of the 4th story. Column-by-column shear demand to capacity ratios, ductility demands, and normalized dissipated hysteretic energy ratios were computed. The proposed framework correctly estimated linear behavior and no damage during the Big Bear earthquake and identified the location of major damage in the beam/column joints located at the fourth floor of the south frame during the Northridge earthquake. The damage indices were identified near unity and above (which corresponds to total failure of the member) in columns with severe damages (wide shear cracks equal or greater than 5 $cm$); between 0.35 and 0.70 in columns with moderate damage (shear cracks smaller than 1 $cm$); and smaller than 0.50 in the remaining columns which did not experienced visible cracks. To the best knowledge of authors, the results presented in this paper constitute the most accurate and the highest resolution damage estimates obtained for the Van Nuys hotel testbed.
	
	
	\section{Data Availability Statement}
	Some or all data, models, or code that support the findings of this study are available from the corresponding author upon reasonable request.
	
	\section{Acknowledgement}
	Support for this research provided, in part, by award No. 1453502 from the National Science Foundation is gratefully acknowledged.
	
	\bibliography{ascexmpl-new}    
\end{document}